# Observation of all-optical bump-on-tail instability
Dmitry V. Dylov and Jason W. Fleischer

**Abstract:** We demonstrate an all-optical bump-on-tail instability by considering the nonlinear interaction of two partially-coherent spatial beams. For weak wave coupling, we observe momentum transfer with no variation in intensity. For strong wave coupling, modulations appear in intensity and evidence appears for wave (Langmuir) collapse at large scales. Borrowing plasma language, these limits represent regimes of weak and strong spatial optical turbulence. In both limits, the *internal* spectral energy redistribution is observed by recording and reconstructing a hologram of the evolving dynamics. The results are universal and can appear in any wave-kinetic system with short-wave/long-wave coupling.

Dynamical instabilities occur in every nonlinear wave system. Perhaps the simplest is modulation instability (MI), in which perturbations grow at the expense of a uniform background. For example, a plane wave propagating in a nonlinear medium will break up into stripes, with a characteristic period determined by a balance between diffraction/dispersion and self-focusing. If the background is statistical, e.g. thermal, then attempts at growth are de-phased by the background, and there is a nonlinear threshold for instability [1-4]. Put another way, mode coupling must be sufficiently strong to generate enough correlation for unstable growth. Once instability begins, the evolution is again characterized by a direct transfer between the background and a preferred scale (this time determined by the correlation length). Here, we consider a different, gradient-driven, instability which couples modes across a range of scales. We show that instability occurs whenever higher-momentum modes are more populated than lower ones, regardless of nonlinear coupling strength, and derive analytic dispersion relations for the growth rates. Experimentally, we examine a bump-on-tail geometry, well-known from plasma physics [5,6], by considering the dynamical coupling of *two* partially-coherent beams in a self-focusing photorefractive medium. For weak nonlinear interactions, the result is momentum (**k**) transfer without any observable variation in intensity (**x**). For strong interactions, both **x**-space modulations and **k**-space dynamics appear. Due to cross-beam coupling, the modulation threshold is higher than the single-beam case, and the spectral form is different. Using wave-kinetic theory, the two-stream dynamics is interpreted as the resonant interaction of speckled light with interaction waves. The results generalize plasma intuition to other fields, allow observation of dynamics difficult to observe in material plasma, and open new pathways toward nonlinear control of statistical light.

We are interested in the spatial propagation of statistical light and its phase-space redistribution due to nonlinearity. Accordingly, we describe the system using a wave-kinetic transport equation [7-9],

$$\frac{\partial f}{\partial z} + \beta \mathbf{k} \cdot \frac{\partial f}{\partial \mathbf{r}} + \frac{\partial \langle \Delta n(I) \rangle}{\partial \mathbf{r}} \cdot \frac{\partial f}{\partial \mathbf{k}} = 0, \tag{1}$$

where $f(\mathbf{r},\mathbf{k},z) = \frac{1}{2\pi} \int_{-\infty}^{+\infty} d\boldsymbol{\zeta} \langle \psi^*(\mathbf{r}+\boldsymbol{\zeta}/2, z) \cdot \psi(\mathbf{r}-\boldsymbol{\zeta}/2, z) \rangle e^{i\mathbf{k}\boldsymbol{\zeta}}$ is the phase-space (Wigner) distribution of the light, $\psi$ is the slowly-varying amplitude, $\beta = \lambda/2\pi n_0$ is the diffraction coefficient, $n_0$ is the base index of refraction, and $\Delta n$ is the nonlinear index change induced by the local light intensity $I = I(x,y,z) = \int f(\mathbf{k}) d\mathbf{k}$. We consider an inertial (slowly-responding) medium in which the nonlinear index change $\langle \Delta n \rangle = \Delta n(\langle I \rangle_\tau)$, where the brackets denote a time average over fast phase fluctuations (to prevent instantaneous filamentation due to intensity hot spots). Equation (1) is essentially a radiation transfer equation [8], valid for slow (long-wavelength) variations in the refractive index when the average speckle size (correlation length $l_c$) of the light is smaller than the beam envelope. That is, Eq. (1) represents a particular example of short-wave/long-wave interactions in a nonlinear medium. Note that the usual form of this dynamics is coupled but has been reduced to a single equation by implicitly absorbing intensity fluctuations in the average index $\langle \Delta n \rangle$.

Speckle dynamics can be considered perturbatively by linearizing Eq. (1) around a background distribution. For simplicity, we study the stability of a homogeneous background in an inertial Kerr medium, with $\Delta n = \gamma \langle I \rangle_\tau$ (the case for the photorefractive nonlinearity, as used in the experiments, is similar). Writing $f(x, k_x, z) = f_0(k_x) + f_1 \exp[i(\alpha x - gz)]$ gives the dispersion relation

$$1 = -\frac{\gamma}{\beta} \int_{-\infty}^{\infty} dk_x \frac{\partial f_0 / \partial k_x}{k_x - g/\alpha\beta}, \qquad (2)$$

where $g = g_R + i g_I$ is the propagation constant for a perturbation mode with wavenumber $\alpha$. The similarity of Eq. (2) to the Vlasov equation from plasma physics prompted Hall *et al.* to interpret the suppression of incoherent MI below threshold as a type of Landau damping [9]. To see this explicitly, we deviate from [9] and consider a Gaussian (vs. Lorentzian) distribution function $f_0(k_x) = (2\pi \Delta k^2)^{-1/2} I_0 \exp(-k_x^2 / 2\Delta k^2)$, where $\Delta k = 2\pi/l_c$. For weak growth $|g_I| \ll |g_R|$ and long wavelengths $\alpha \ll g/(\beta k_x)$, expanding the denominator of Eq. (2) gives

$$g_R^2 = g_P^2 \left(1 + 3\alpha^2 \lambda_D^2\right) \qquad g_I \cong \frac{\pi}{2} \gamma \alpha \sqrt{\frac{\gamma I_0}{\beta}} \cdot \left(\frac{\partial f_0}{\partial k_x}\right)\bigg|_{k_x = g/\alpha\beta} \qquad (3)$$

where $g_P = \sqrt{\gamma \beta I_0}$ is an effective plasma frequency and $\lambda_D = \beta \Delta k / g_P$ is an effective Debye length. Equation (3) is a Bohm-Gross dispersion relation [5] for nonlinear statistical light, showing that optical speckles can interact via Langmuir-type waves. Growth or damping of these waves is a resonant process that depends on the relative (spatial) phase velocity of the underlying quasiparticles (speckles). From the expression for $g_I$, it is clear that there are no growing modes if $\partial f_0 / \partial k_x < 0$, e.g. for a quasi-thermal Gaussian distribution, since on average more quasiparticles travel slower than the interaction wave than faster. However, the weak limit used to derive Eq. (3) breaks down when $\alpha\lambda_D \approx 1$, or $\gamma I_0 \approx \beta\langle\Delta k^2\rangle$; in this case, the growth rate exceeds the rate of statistical de-phasing (spectral bandwidth) of the background, causing intensity modulations to appear [1,6]. Interestingly, this strong-coupling condition becomes an exact instability threshold when a Lorentzian distribution is used in Eq. (2) [3]. Not considered before, however, was the possibility for optical instability by *inverse* Landau damping when $\partial f_0/\partial k_x > 0$. A prime example is the "bump-on-tail" (BOT) instability, well-known from plasma physics, in which a non-equilibrium hump is added to one side of an equilibrium distribution. However, it should be clear from the above derivation that BOT dynamics should occur in any wave-kinetic system, including hydrodynamics [10], optics, and (potentially) Bose-Einstein condensates. To our knowledge, the BOT instability has never before been demonstrated outside of a plasma context.

Here, we observe this instability all-optically by considering the nonlinear interaction of two partially-coherent spatial beams. The experimental setup is shown in Fig. 1. A statistical light input is created by focusing light from a 532nm CW laser onto a ground-glass diffuser and then imaging into a photorefractive SBN:60 ($Sr_{0.6}Ba_{0.4}Nb_2O_6$) crystal. The correlation length, and correspondingly the spectral bandwidth, can be changed by varying the magnification properties of the imaging lens [11]. To create a bump-on-tail distribution (Fig. 1 inset), the spatially-incoherent beam is split using a Mach-Zehnder interferometer, attenuated in one arm, and then recombined on the input face of the crystal. For SBN, the nonlinear index change $\Delta n = \kappa E_{app}\langle I\rangle/(1+\langle I\rangle)$, where $E_{app}$ is an electric field applied across the crystalline c-axis and $\kappa = n_0 r_{33}(1+\langle I_0\rangle)$ is a constant depending on the base index of refraction $n_0$, the electro-optic coefficient $r_{33}$, and the spatially-homogeneous incident light intensity $\langle I_0 \rangle$ [12,13]. In the experiments, the beams have a relative angle of 0.3°, the intensity ratio is fixed at 3:2, and the strength of the nonlinearity (wave coupling) is controlled by varying the applied voltage (similar results occur at other angles and intensities). To observe the interaction, light exiting the crystal is directly imaged in both position (**x**) space and momentum (**k**) space, the latter by performing an optical Fourier transform.

For comparison and calibration, we performed a single-beam MI experiment with the main $\bar{k}_x = 0$ hump (not shown). In this case, the background distribution is Gaussian with a correlation length $l_c$ = 176 μm, and no intensity modulations appeared until the voltage reached 0.9 kV. Using $n_0$ = 2.3 and $r_{33}$ = 255 pm/V, this corresponds to a nonlinear index change of $\Delta n = 8 \cdot 10^{-4}$. Above this threshold, two symmetric momentum peaks appear at $k_x/k = \pm 5.6 \cdot 10^{-3}$. This is the same behavior as in [4] but quantitatively calibrated to our initial input conditions and particular crystal.

All-optical examples of a wave-kinetic "bump-on-tail" instability are shown in Figs. 2-4. Fig. 2 shows the behavior for weak interaction. In this case, the photorefractive nonlinearity is turned on by applying a 0.7 kV voltage bias across the crystal, *below* the 0.9 kV bias necessary to trigger single-beam MI. As shown in Figs. 2(c,f), nonlinear modes are excited precisely in the expected region of positive slope, growing until there is no more driving gradient (a process known as quasilinear flattening [5,6]). Remarkably, the momentum-space distribution is changed [Figs. 2(e,f)] while the position-space intensity shows no observable variations [Fig. 2(d)].

The spectral dynamics depends on the statistics of the interacting beams (Fig. 3). Local correlation measurements can reveal details of the speckle-wave coupling, but a simpler measure can be obtained from the visibility $v = [f(k_1) - f(k_{01})]/[f(k_1) + f(k_{01})]$ of the angled hump, as shown in Fig. 3c. The efficiency of the flattening depends on the relative difference between the input (linear) spectrum [Fig. 3(a)] and the output spectrum [Fig. 3(b)]. Taking the relative gain of the unstable modes (versus that of the $\bar{k}_x = 0$ peak) as a measure of efficiency, $\eta = [f^{NL}(k_{01}) - f^{Lin}(k_{01})]/[f^{NL}(k_0) - f^{Lin}(k_0)]$, Fig. 3(d) shows interaction behavior that is relatively insensitive to nonlinear coupling strength but highly sensitive to beam statistics. If the beam is too incoherent, then attempts at spectral energy transfer are de-phased. If the beam is too coherent, then the system loses its statistical nature (and thus its wave-kinetic properties). More rigorously, the first condition states that the angular separation between the beams must be greater than the spectral width of the distribution, while the second condition states that if the relative bandwidth is too small, then there are too few quasiparticles (speckles) in resonance with the growing waves [5,6]. As a result, there is an optimal correlation length, for a given intensity ratio and angle, for efficient dynamical coupling.

For stronger nonlinearity, the system enters a regime of strong wave coupling, significantly distorting the original distribution in **k**-space and creating modulations in **x**-space (Fig. 4). These modulations are different from those arising from MI, however, as the spectrum in Figs. 4b,c shows a range of modal excitation (between the original humps), rather than the symmetric high-$k$ side lobes characteristic of MI (e.g. [3,14]). Using our reference correlation length $l_c = 176$ $\mu m$, as in Fig. 2, we observe that the required nonlinearity for modulations is 1.1 kV, stronger than the one needed for single-beam MI. That is, the presence of a second statistical beam further suppresses the growth of modulations. Moreover, the appearance of modulations coincides with a breakdown of the quasilinear plateau and a resumption of wave growth in the unstable, non-equilibrium region [Fig. 4(c)].

The higher threshold can be understood by returning to the strong-coupling condition $\gamma I_0 \approx \beta \langle \Delta k^2 \rangle$ obtained from Eq. (3). It is clear that for a given spectral width, additional intensity lowers the required value of $\gamma$ for instability [3,15]. However, the presence of a second beam increases the effective bandwidth due to cross-beam interaction. If this spectral spread is more than the increase in intensity, a higher value of $\gamma$ would be required. A simple estimate can be obtained by considering the variance of two Gaussian beams $\exp(-k_x^2/\Delta k^2) + A\exp(-(k_x - \delta k)^2/\Delta k^2)$, which is $\Delta k^2 + \delta k^2 \cdot A/(1+A)^2$. For $A = 2/3$ and $\delta k \sim \Delta k$, as in the experiments, there is an *increase* in threshold nonlinearity from $\gamma$ to $(31/25)\gamma$. Given the measured single-beam MI threshold of 0.9 kV, the predicted double-beam threshold of 1.12 kV matches the observed value.

The different behaviors above and below the modulation threshold are the result of different nonlinear dynamics within the initial distribution. Experimentally, we can observe this by taking advantage of the slow photorefractive response time of SBN and recording a volume hologram of the interactions. Subsequently, we can block one of the beams and use the other as a probe of the coupling, observing the energy transfer that would have happened if the other beam were present [16]. These holographic reconstructions are shown in Figs. 2(c,f), and 4(c). For linear propagation [Fig. 2(c)], each beam maintains its Gaussian form, as there is no nonlinear intensity interaction to induce an index change. By contrast, there are significant changes in the nonlinear cases. For weak coupling [Fig. 2(f)], light originally in the perturbative bump (shown in red) is seen to flow towards lower momentum states, while light from the "equilibrium" distribution (shown in blue) scatters in the opposite direction. For strong coupling [Fig. 4(c)], the momentum transfer is *asymmetric*. The "thermal" light is unchanged, while the "nonthermal" distribution looks *bimodal*, with half the intensity in the original angled hump and half centered at $\bar{k}_x = 0$, beyond the initial instability range of positive slope.

At this point, it is useful to revisit the plasma correspondence and interpret the scattering dynamics from a quasiparticle (speckle) perspective. From this viewpoint, the instability mechanism is essentially a resonant process, in which small-scale wavepackets generate and interact with large-scale modulations [17]. The coupling threshold $\gamma I_0 \approx \beta \langle \Delta k^2 \rangle$ then separates the dynamics between regimes of weak and strong spatial turbulence. Indeed, weak (quasilinear) turbulence theory in plasma is *characterized* by the formation of a **k**-space plateau and the bidirectional transfer of momentum between the thermal and nonthermal distributions [5,18]. In the theory of strong turbulence, the "thermal," nonresonant distribution is unchanged but the resonant distribution is greatly affected by wave-wave interactions [6]. In this case, there should be a direct transfer of momentum towards large scales ($\bar{k}_x = 0$), a stimulated scattering process known as Langmuir condensation in plasma physics [6,19]. All of this is consistent with the observations in Figs. 2f and 4c. Coupled with the correspondences in Eqs. (2) and (3), these results strongly suggest that the nonlinear propagation of statistical light can be considered as a quasi-plasma of interacting speckles. As such, there is clearly much potential in controlling correlation dynamics and optical energy distributions based on plasma-type wave phenomena. It also suggests that many plasma phenomena that are difficult to observe in material systems, due to screening effects, can be measured more easily in their photonic analogues.

In conclusion, we have observed basic wave-kinetic instabilities and bump-on-tail-driven turbulence using statistical light. To our knowledge, this is the first demonstration of a bump-on-tail instability outside of a plasma context. The results lay the foundation for all-optical studies of plasma physics and hold potential for correlation-based photonic devices.

## ACKNOWLEDGMENTS

We thank P.H. Diamond, N.J. Fisch, and H. Buljan for very valuable discussions. This work was supported by the NSF and AFOSR.

## REFERENCES

[1] A. A. Vedenov and L. I. Rudakov, Doklady Akademii Nauk SSSR **159**, 767 (1964) [Sov. Phys. Dokl. **9**, 1073 (1965)].
[2] A. A. Vedenov, A. V. Gordeev and L. I. Rudakov, Plasma Physics **9**, 719 (1967).
[3] M. Soljacic *et al.*, Phys. Rev. Lett. **84**, 467-470 (2000).
[4] D. Kip *et al.*, Science **290**, 495-498 (2000).
[5] N. A. Krall and A. W. Trivelpiece, Principles of Plasma Physics (McGraw-Hill, 1973).
[6] M. V. Goldman, Rev. Mod. Phys. **56**, 709 (1984).
[7] D.N. Christodoulides, T.H. Coskun, M. Mitchell, and M. Segev., Phys. Rev. Lett. **78**, 646 (1997).
[8] V.V. Shkunov and D.Z. Anderson, Phys. Rev. Lett. 81, 2683-2686 (1998).
[9] B. Hall *et al.*, Phys. Rev. E **65**, 035602 (2002).
[10] G. E. Vekstein, American Journal of Physics **66**, 886 (1998).
[11] W. Martienssen and E. Spiller, Am. J. of Phys. **32**, 919 (1964).
[12] M. Segev *et al.*, Phys. Rev. Lett. **73**, 3211-3214 (1994).
[13] D. N. Christodoulides and M. I. Carvalho, J. Opt. Soc. Am. **B** − Opt. Phys. **12**, 1628-1633 (1995).
[14] Z. Chen, J. Klinger, and D.N. Christodoulides, PRE **66**, 066601 (2002).
[15] L. Helczynski *et al*. IEEE J. Sel. Top. Q. Electr. **8**, 408-412 (2002).
[16] C. Anastassiou *et al*., Phys. Rev. Lett. **83**, 2332-2335 (1999).
[17] B.B. Kadomtsev, Plasma Turbulence (Academic Press, New York, 1965).
[18] A. N. Kaufman, J.Plasma Phys. **8**, 1 (1972).
[19] P. A. Robinson, Rev. Mod. Phys. **69**, 507-573 (1997).

**FIGURE CAPTIONS**

**Fig. 1.** Experimental setup for observation of bump-on-tail instability. Light from a 532nm CW laser is made statistical by first passing it through a ground-glass diffuser and then imaging it into a nonlinear SBN:60 crystal. The correlation length of the light is adjusted by changing the image magnification of lenses $L1$ and $L2$, the double-hump distribution is created using a Mach-Zehnder interferometer (mirrors $M$ and beamsplitters $BS$), and the strength of nonlinear wave coupling is adjusted by varying an applied voltage $V$ across the SBN crystal. Light exiting the crystal is directly imaged in both position (**x**) space (in digital camera DC1) and momentum (**k**) space (in DC2 by performing an optical Fourier transform of the $L3$ focal plane FP).

**Fig. 2.** Experimental results of all-optical bump-on-tail instability. Left column: crystal's exit face after linear propagation (no applied voltage); right column: exit face after nonlinear propagation (applied voltage of 0.7 kV, in the weak-coupling regime). (a,d) Intensity in position (**x**) space; (b-f) power spectrum in momentum (**k**) space. The blue and red curves in (c,f) show holographic readouts of single-beam propagation for the straight (blue) and angled (red) distributions, respectively. The linear case (c) shows no change in profiles, as there is no interaction between the beams. By contrast, the nonlinear case (f) shows counterpropagating momentum flow, resulting in quasilinear flattening of the overall distribution.

**Fig. 3.** Dynamical coupling as a function of correlation length and nonlinearity. (a,b) Power spectra at 0 kV (linear) and 0.7 kV, respectively. (c) Visibility of the angled hump. (d) "Efficiency" of nonlinear flattening (the dotted line is a guide for the eye). Curves and bars in (a,c) are numbered for correlation lengths of 243 $\mu$m(1), 206 $\mu$m (2), 176 $\mu$m (3), 142 $\mu$m (4), and 109 $\mu$m (5).

**Fig. 4.** Spatial optical turbulence in the strong-coupling regime. (a) Position-space intensity modulations at 1.6 kV. (b,c) Power spectrum, showing significant k-space distortion to single-humped profile. Holographic reconstruction (c) shows no change in nonresonant distribution (blue) but large momentum transfer in resonant distribution (red). The bimodal form of this transfer is suggestive of wave (Langmuir) condensation.

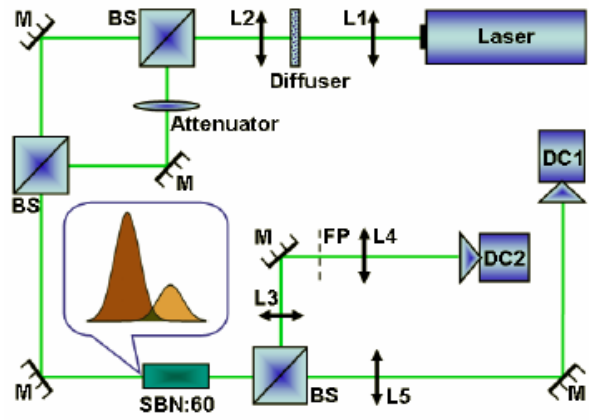

**FIGURE 1**

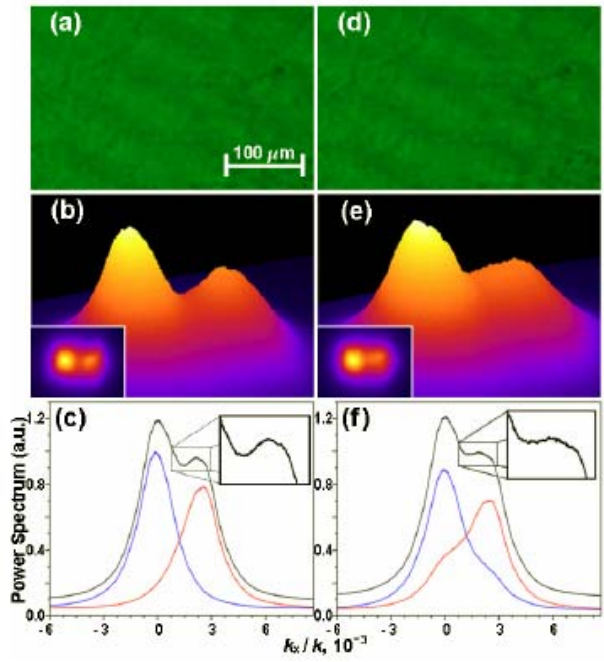

**FIGURE 2**

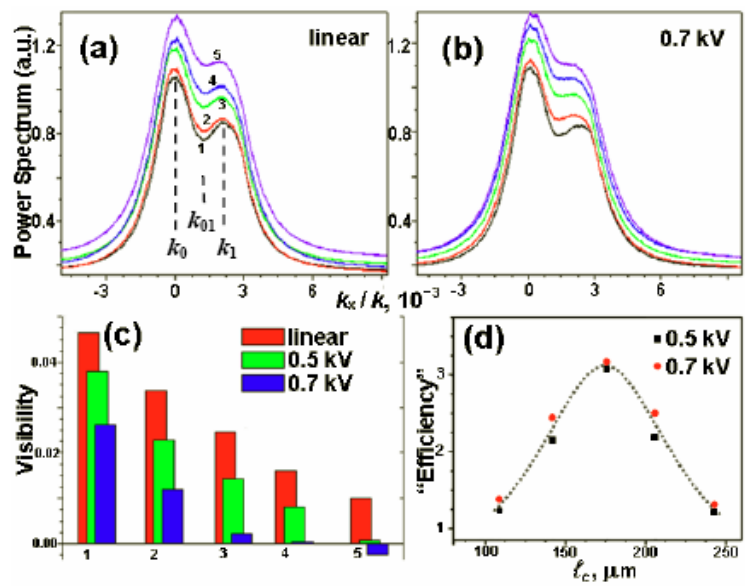

**FIGURE 3**

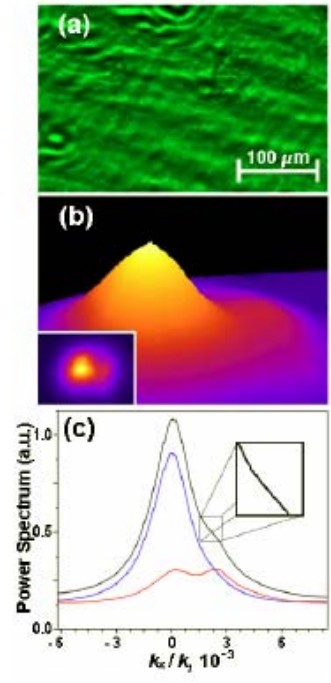

**FIGURE 4**